# Closed Eyes and Coil Size – Effects on Motor Threshold and Intracortical Inhibition, measured with TMS

Meher Sabharwal BA (Hons)[1*], Narin Suleyman MPhil.[1,2*], Gabriel R. Palma PhD[1], Roisin McMackin PhD[1]

[1]Discipline of Physiology, School of Medicine, Trinity College Dublin, the University of Dublin, Ireland

[2]Department of Neurology, Beaumont Hospital, Dublin, Ireland

*Joint first authors

**Corresponding author:** Dr Roisin McMackin. Address: Discipline of Physiology, Trinity Biomedical Sciences Inst., 152-160 Pearse St., Dublin, Ireland. D02R590. Phone: + 3531896 2718

**Abstract**

*Rationale* Transcranial magnetic stimulation (TMS)-based measures such as resting motor threshold (RMT) and short interval intracortical inhibition (SICI) are widely employed to study motor cortical and corticospinal tract function, and effects of diseases and drug therapies thereon. However, the effect of key experimental factors, including as eye state (open or closed) or stimulating coil size, remain unclear. As such, it is unknown whether these factors must be kept consistent across multi-center studies, and whether differences in such factors may underpin contradictory findings in existing literature.

*Materials and Methods* Threshold tracking TMS was employed to measure RMT and SICI (3ms interstimulus interval, conditioning at 70% of RMT) in 21 alert and awake, healthy controls. Motor evoked potentials were recorded from abductor pollicis brevis. Both RMT and SICI were measured under 6 conditions, while eyes were open or closed, using 3 figure-of-eight coils of differing winding diameter. Mixed effects modelling was employed to investigate effects of eye state and coil size on each measure.

*Results* RMT was found to be significantly higher for the smallest (30BFT) coil compared to both larger (50BFT and 70BF) coils. No difference in SICI was identified across coil sizes. Eye state was not found to affect either RMT or SICI measurements.

*Conclusions* Measurements of RMT and SICI can be considered comparable if recorded with eyes open or closed, provided the individual is awake and alert. Measurements of SICI recorded with figure-of-eight coils of different size can be considered comparable.

**INTRODUCTION**

Transcranial magnetic stimulation (TMS) is a powerful noninvasive tool with both diagnostic and therapeutic applications in neurological disorders. Single- and paired-pulse TMS measures in particular are gaining traction as biomarkers of some neurodegenerative diseases, in addition to their utility for understanding health and disease-related neurophysiology[1].

The resting motor threshold (RMT) is defined as the minimum stimulus intensity required to generate a motor-evoked potential (MEP) of at least 50μV in 50% of trials. The RMT is a widely used measure to study overall corticospinal tract excitability and to standardise TMS protocols and measurements across individuals and across longitudinal recordings within individuals[2]. Paired pulse measures provide additional insights into cortical circuits which regulate corticospinal tract excitability. For example, short interval intracortical inhibition (SICI) recorded with an interstimulus interval (ISI) of approximately 2-3ms is widely used to study motor GABA-Aergic inhibition and the effects of neurodegenerative and psychiatric disorders thereon[1, 3]. Further, SICI has consistently been shown to be lower in people with MND compared to controls and mimic conditions[4-6]. As a result, this measure is now used as a supportive biomarker in ALS diagnosis[7], reflecting increasing application of such measures across clinical settings as well as in multi-centre clinical trials.

In order to maximise the reliability of these measures within participants, and to avoid effects of lurking covariates across research sites and longitudinal measurements, participant state (e.g. sitting/lying relaxed with eyes open) and recording equipment protocols (e.g. coil size and shape) are typically kept consistent within studies, and have been speculated as the reason for differences between findings across studies[8-10]. However, the effects of such lurking variables are rarely systematically tested. As such, for many controlled factors in TMS-based studies, their effect, if any, on measurements of interests remains unclear.

One factor that is typically controlled during recording of TMS-based measures of motor network function is eye state. Namely, studies typically report that volunteers are asked to keep their eyes open[8, 11, 12]. This requirement may be implemented under the assumption that having eyes closed may result in the individual falling asleep, where non-REM sleep is associated with increase in RMT and increased SICI, compared to wakefulness[13-15]. However, very few studies have tested the effects of having eyes open or closed on RMT and SICI independently of sleep-related effects, in awake participants. Notably, these few studies have reported either greater MEP amplitudes[16], or no difference[9, 17, 18], during eyes closed recordings compared to those with eyes open. Where greater MEP recruitment was found in the eyes closed state (in a lit room), a similar effect was observed if eyes were kept open under blackout goggles, preventing visual stimulation[16]. Only one of these studies investigated SICI, where no effect of eye state was reported[18].

Another factor that is typically controlled within studies and across study sites is the size of the TMS coil utilised, as larger coils induce currents of greater depth and lesser focality at a given stimulation intensity[19, 20]. Correspondingly, it is assumed that stimulation with a larger coil will result in lower RMT measurements, due to larger figure-of-eight coils. However, while the effect of coil shape, such as figure-of-eight coils and circular coils, on SICI has been repeatedly investigated[21, 22], no study to date has examined if coil size alone affects this measure. As such, it remains unclear if it is necessary for all research/clinical sites to use the same coil size to record comparable SICI measurements, or if differences in coil size between studies might explain differences in findings.

In this study, we aimed to systematically investigate the influence of eye state (open or closed) and figure-of-eight coil size (30, 50 or 70mm outer winding diameter) on RMT and SICI in awake individuals. By addressing these questions, this study seeks to more definitively determine the nature and magnitude of any effects of these variables on these measures, and if these factors should be kept consistent across recordings, individuals, or sites, within TMS-based research or clinical applications.

**METHODS**

*Ethical approval*

Ethical approval was obtained from the research ethics committees of St James Hospital (REC ref: 2017–02) and the Trinity College Dublin School of Medicine (REC ref: 3538). All participants were over 18 years of age and provided informed written consent prior to participation. All work was performed in accordance with the Declaration of Helsinki.

*Participants*

Inclusion criteria were as follows: Aged 18 years old or over with RMT ≤99% of maximum stimulator output (%MSO) when recorded with a 50BFT coil. Exclusion criteria included any safety-related contraindications to TMS according to international consensus guidelines[23-25], in addition to a current diagnosis of neurological or neuropsychiatric conditions, use of medications known to affect the nervous or muscular system, any known structural or functional nervous system abnormalities and any prior history of seizure disorders. Participants were also selected to achieve similar participation of biological males and females of similar ages. Handedness was assessed by the Edinburgh Handedness Inventory[26]. All participants reported they had a typical night of sleep prior to attending and were confirmed to be awake throughout the recording session. Regular breaks were taken between measures to offset fatigue and maintain alertness.

*Transcranial magnetic stimulation*

Data were recorded in the Clinical Research Facility, St James Hospital, Dublin and at Trinity Biomedical Sciences Institute, Dublin. Participants were seated with arms resting either on arm rests or in the participant's lap, with elbows bent at a 90-120° angle. Monophasic magnetic stimuli were delivered via a DuoMag MP dual stimulator (Deymed Diagnostics s.r.o., Hronov, Czech Republic). Three coils were employed during this study, namely the DuoMag 30BFT, 50BFT and 70BF figure-of-eight shaped coils (Deymed Diagnostics s.r.o., Hronov, Czech Republic), which have external winding diameters of 30, 50 and 70mm respectively.

Pulses were delivered over the left hemisphere to the scalp hotspot for the right abductor pollicis brevis (APB). Coils were positioned such that magnetic pulses induced a flow of electrical current directed perpendicularly across the precentral gyrus, proceeding from posterior to anterior (PA). The hotspot was identified as the position on the scalp at which reliable MEPs could be evoked in the target muscle with the lowest stimulation intensity. Consistent coil placement was ensured by marking landmarks which corresponded with markings on the back of the coil, on a cloth cap secured to the head (as described in detail in McMackin et al. 2024[4]). The procedure of identifying the hotspot was completed for each coil to ensure all stimuli were delivered to the point at which the least stimulation intensity was required to evoke a response in the target muscle. Hotspotting was repeated for each coil to ensure accurate landmarks were noted on the cap for each coil. No substantial differences (<1cm) in hotspot location were identified across coils. Electromyography (EMG) was recorded from the abductor pollicis brevis (APB) of the dominant hand. MEPs were recorded using single channel surface EMG in a belly-tendon montage, using pairs of Ag-AgCl electrodes placed approximately 2cm apart and a reference electrode placed on the ulnar head of the right wrist. EMG signals were amplified (gain=1000) and band-pass filtered (10–1000 Hz) using BioPac EMG100C amplifiers (Biopac Systems UK, Pershore, UK). Surrounding environmental electrical noise was subtracted after amplifying the signal using either a Humbug single channel or D400 multichannel noise eliminator (Digitimer Ltd., Welwyn Garden City, UK). Signals were digitized at a sampling rate of 10 kHz (Micro1401, CED, Cambridge, UK), and recorded using Signal software (Signal 7.01, CED, Cambridge, UK).

*Experimental paradigm*

RMT and SICI were measured for all combinations of coil sizes (30BFT, 50BFT and 70BF) and eye states (open or closed). During "eyes open" (EO) recordings, participants were instructed to look straight ahead at a blank wall. During "eyes closed" (EC) recordings, participants were instructed to close their eyes but to remain awake. Participants were monitored to ensure they remained awake throughout the session. In all cases, participants were advised to relax and let their mind wander, and not to specifically attend to the ongoing TMS. Measurements for each condition were collected in pseudorandomized order, while ensuring that for each SICI measure, the RMT measure required to calculate conditioning intensity had already been recorded. As such, participants did not keep their eyes closed for longer than 3 minutes at any point during recording. All recordings were undertaken in a lit room with consistent lighting across measurement recordings.

Both RMT and SICI were measured under a threshold tracking procedure, using a fully automated maximum likelihood parameter estimation by sequential testing (PEST) protocol[27], automated using Signal (Version 7.01, CED Ltd., Cambridge UK) and MATLAB (R2016a, MathWorks Inc., MA, USA) scripts, described in detail in McMackin et al., 2024[4]. In brief, this procedure utilises a sigmoid-shaped logistic function to determine the stimulation intensity (motor threshold) at which there exists a 50% probability of eliciting an MEP with peak-to-peak amplitude greater than a defined value, which was 50µV in the case of this study for all measures. Two predefining datapoints were input to the function, one for a stimulation intensity which consistently did not evoke MEPs above the target amplitude, and one for a stimulation intensity 1.5 times this value which consistently did evoke MEPs above target. Such predefined datapoints guide initiation of threshold tracking to within this stimulation intensity range. Thereafter, 18 datapoints (delivered stimulus intensity and whether MEP peak-to-peak amplitude exceeded target) were successively input to the PEST algorithm to estimate threshold within 95%-105% of true threshold.

Identical procedures were employed to estimate RMT and the "conditioned threshold target" (CTT), with the exception that in the case of CTT, each tracked "test" stimulus was preceded by a "conditioning" stimulus applied at 70% of the RMT recorded for the same coil and eye state. The conditioning stimulus was delivered 3ms prior to the test stimulus. After data collection, $SICI_{3ms}$ was then calculated as follows:

$$\% \text{ of Inhibition or Facilitation } = \left(\frac{CTT - RMT}{RMT} \times 100\%\right)$$

Where positive values reflect inhibitory effects of conditioning, and negative values reflect facilitatory effects. Trials with root mean squared (RMS) EMG amplitude greater than 20μV in the 200ms immediately prior to stimulation were rejected and repeated.

*Statistical analysis*

A preliminary investigation of differences in mean baseline RMS-EMG across conditions was performed to determine if this factor may reflect a pertinent lurking variable. Mean baseline RMS-EMG was compared by repeated measures ANOVA and Tukey's honestly significant difference post-hoc testing for those where measurements could be collected under all 6 conditions (n=18 for RMT, n=13 for SICI). These analyses revealed significant differences in baseline activity while RMT and SICI were recorded with different coils (reported in supplementary results).

As such, models were employed to simultaneously examine the effects of coil size, eye state and mean baseline RMS-EMG on RMT and SICI. A beta Generalised Additive Model for Location, Scale and Shape (GAMLSS) [28] was fitted to the RMT data, while a Linear Mixed Effects Model was fitted to the SICI data. Model goodness-of-fit was assessed via worm plots [28, 29]. The choice of a beta GAMLSS is based on the nature of the RMT response variable, which is bounded between 0 and 100%, and comparisons based on Akaike Information Criterion (AIC), Bayesian Information Criterion (BIC) and visual inspection of goodness-of-fit, which demonstrate that this distribution is more adequate for the presented dataset than alternative distributions. By contrast, SICI can span both positive and negative values, and a normal distribution-based model was found to provide similar or lower AIC and BIC to alternative potentially suitable distributions for the presented dataset.

Likelihood-ratio (LR) tests were used to assess the significance of the fixed effects and exclude unnecessary effects from the final reported model. For both models, LR tests indicated that baseline RMS-EMG and interaction effects had no significant effect on RMT or SICI. As such, only the effects of eye state and coil size were included in the linear predictor as a fixed effect, with subject included as a random effect for the mean. For GAMLSS, only the fixed effects of eye state and coil size were included in the linear predictor for dispersion. Fixed effects were deemed significant for LR test statistic (LRT) associated p-values<0.05.

All analyses were carried out using R[30]. The nature of any significant fixed effect was examined through post hoc comparison of overlap in model-derived confidence intervals for measure estimates for each condition of that effect.

Analysis of the effects of eye state and coil size via repeated measurement ANOVA (excluding participant data where measures could not be recorded for all 6 conditions) provided the same qualitative conclusions. However, we have opted to report findings based on the aforementioned modelling, as model selection criteria and the goodness-of-fit assessment demonstrated that a beta GAMLSS is more suitable for the presented RMT dataset. In addition, the use of modelling facilitated the inclusion of all data collected.

**Results**

*Participant cohort*

A total of 21 people (age median [range]=22 [18-50] years), consisting of 11 women and 10 men, were included in the study. All but two participants were right-handed. Male and female ages were not significantly different, as determined by Mann Whitney U testing (p=0.47). Notably, one participant had an outlying age of 50 years, with age range excluding this participant being 18-27 years. Exclusion of this participant and repetition of all reported analyses did not affect any of the findings of this study.

In 3 people, RMT was not recorded for all 6 experimental conditions. Specifically, for one individual, the 70BF coil was unavailable at the time of recording. For another, RMT exceeded 100% MSO when using a 30BFT coil for both eye states, and for the third relaxation was insufficient during recording with EO using 30BFT and 70BF coils. As such, SICI measures were unavailable for these conditions in these individuals. For an additional 5 people, SICI could not be recorded for all 6 conditions. This was a result of CTT exceeding 100% MSO when using a 30BFT coil with EO (n=1) or for both eye states (n=2), or insufficient relaxation during CTT recording using a 30BFT coil with EO (n=1) or closed (n=1).

*Resting motor threshold*

Figure 1 illustrates all recorded RMT measures, paired across eye state (Figure 1a) and coil size (Figure 1b) within individuals. Interaction between eye state and coil size was not significant (DF = 1.99, LR = 0.21, $p$ = 0.90) for RMT measures. Difference in eye state was not found to significantly affect RMT (DF = 1, LR = 2.52, $p$ = 0.11). A significant effect of coil size on RMT was observed (DF = 2.06, LR = 58.41, $p$ < 0.01). Post hoc comparison of each coil size attributed this effect to a greater RMT when using a 30BFT coil than either larger coil, while measures with larger coil sizes were comparable. Repetition of the analysis including only the 18 individuals for whom RMT was recorded for all 6 conditions (age median [range]=22 [18-50] years), 10 women and 8 men) did not change the qualitative findings of this model.

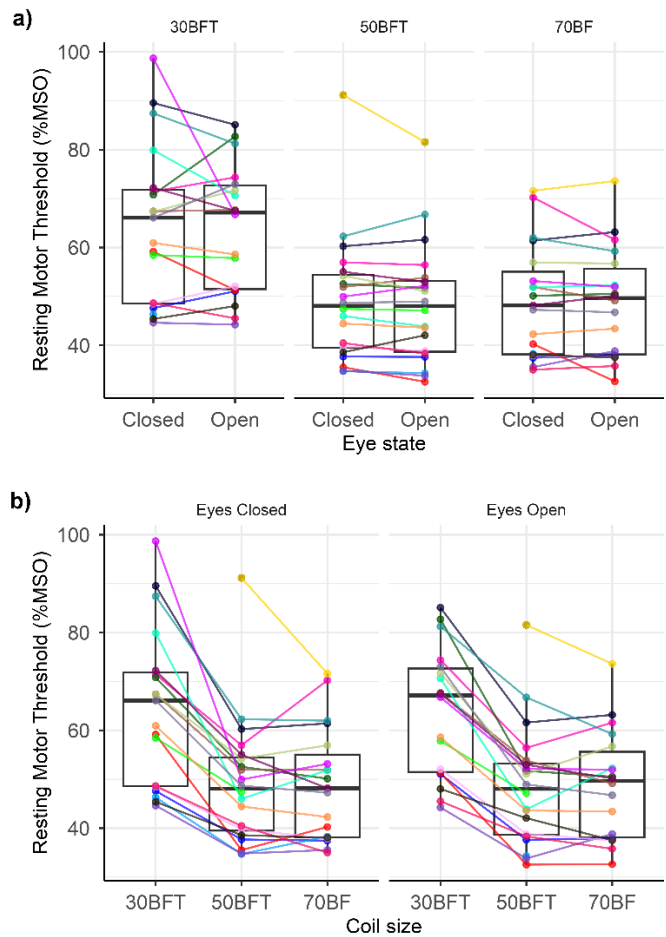

**Figure 1. Comparison of paired resting motor threshold (RMT) values across conditions, with paired measures illustrated between eye states (A) and between coil sizes (B).** Boxes

outline the interquartile range ($IQR$) calculated from the 25th and 75th quartiles, with horizontal lines indicating the median. The upper and lower whiskers are calculated as $\pm 1.5 * IQR$.

*Short interval intracortical inhibition*

Figure 2 illustrates all recorded SICI measures, paired across eye state (Figure 2a) and coil size (Figure 2b) within individuals. Interaction between eye state and coil size was significant (DF = 1.98, LR = 0.22, $p$ = 0.89) for SICI measures. No significant difference in SICI was identified when recorded under different eye states (DF = 1.00, LRT = 0.69, $p$ = 0.40) or coil sizes (DF = 2.21, LRT = 2.59, p = 0.31). Repetition of the analysis including only the 13 individuals for whom SICI was recorded for all 6 conditions (age median [range]=22 [18-50] years, 6 women and 7 men) did not change the qualitative findings of this model.

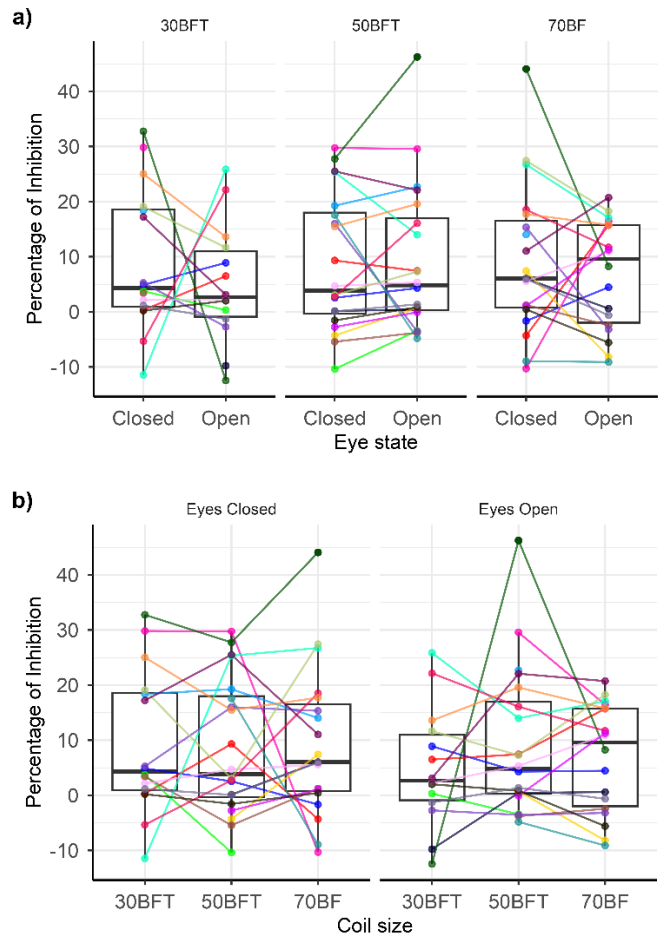

**Figure 2. Comparison of paired short interval intracortical inhibition (SICI) values across conditions, with paired measures illustrated between eye states (a) and between coil sizes (b).** Positive y axis values indicate an inhibitory effect of conditioning, negative y values indicate a facilitatory effect of conditioning. Boxes outline the interquartile range ($IQR$) calculated from the 25th and 75th quartiles, with horizontal lines indicating the median. The upper and lower whiskers are calculated as $\pm 1.5 * IQR$.

**Discussion**

*Coil size*

We have demonstrated a significant effect of coil size on RMT, with post hoc analysis revealing that the 30BFT coil required significantly higher stimulation intensities than both the 50BFT and 70BF coils. No significant SICI-based differences were observed among coil types.

Prior studies have demonstrated that coil geometry and size can influence MEP amplitudes and areas, likely due to differences in the magnitude of descending volleys and subsequent recruitment of lower motor neurons[31]. Similarly, differences in SICI at 2ms ISI have been reported when comparing figure-of-eight coils and circular coils due to differences in magnetic field patterns of both the coils[21].

By simulating the electric field distributions induced by 50 different TMS coils, Deng and colleagues hypothesised a depth-focality trade-off: larger coils generate less focal stimulation but do result in higher depth of penetration of the electric field, as compared to smaller coils[19]. This may explain why smaller coils, such as the 30BFT, necessitate higher stimulation intensities to achieve comparable cortical effects. Our findings are consistent with this hypothesis, indicated by the significantly higher RMT values for 30BFT coils compared to 50BFT and 70BF coils, some even exceeding 100%MSO. The 30BFT coil is seldom used in human studies except in multicoil, multisite studies, likely due to practical limitations such as excessive heating and participant discomfort at high intensities. These issues were similarly encountered in our study and may limit the coil's utility in experimental paradigms requiring high stimulation thresholds.

No coil-based differences were noted among SICI measures despite RMT variations. As SICI is a measure of relative inhibition, it may be less sensitive to absolute intensity changes, supporting its robustness as a marker of intracortical inhibitory processes[32].

*Eye state*

We have demonstrated that there are no significant differences between EO and EC states in RMT and SICI measures across coil sizes. Participants in our study kept their eyes closed only briefly (2mins), in a lit room, and were instructed to relax and let their minds wander during stimulation. Our results are consistent with prior findings of no effect of eye state on MEP amplitude during RMT or SICI in awake individuals, recorded with a 70BF coil[18]. It should be noted that while MEP amplitude is used to index corticospinal recruitment and measure motor circuit excitability, it is subject to variability based on physiological and methodological factors like pre-stimulus muscle contraction, intertrial variability in neuronal excitability, and stimulation intensity[33]. Our study did not examine motor cortex excitability based on raw MEP amplitude. Instead, we employed a threshold-tracking method which provides a more stable measure of cortical excitability by determining the motor threshold based on the likelihood of evoking an MEP at a given target amplitude[34]. As far as we are aware, our study is the first to use threshold tracking in a study of eye state on motor network excitability.

Our findings contradict those of Leon-Sarmiento et al., who found that MEP amplitudes were increased during brief eye closure (specific duration unspecified) in a lit room as well as after more prolonged (30 minute) eye closure[16]. By contrast, Cambieri et al. found no change to the MEP amplitude during very brief periods (2.5s) of eye closure under room lights[17]. These differences may be due to the differences in coil shape and size between the studies. Leon-Sarmiento et al. used a large, 9cm circular coil, while Cambieri et al.'s study, like ours, used a figure-of-eight-shaped coil. The depth/focality ratio of figure-of-eight-shaped coils, might explain this discrepancy.

Regarding alertness or drowsiness, as opposed to eye state, TMS-EEG studies suggest that both spontaneously decreased[37] and EO-state dependent decreased mu/alpha power[38] are associated with larger MEPs in single pulse trials, in line with studies suggesting that changing alertness levels throughout a session[39] and alterations in somatosensory oscillations may affect primary motor cortex excitability[40, 41]. Moreover, fluctuations in neuronal polarization during NREM sleep have been shown to attenuate the motor cortex response to stimulation, with smaller MEP amplitudes and higher intracortical inhibition being observed during periods of hyperpolarization compared to pre-sleep wakefulness[13]. As such, where closure of eyes is confounded by decreased alertness, an increase in motor thresholds may instead be observed. It is unclear how the studies of Leon-Sarmiento et al. and Cambieri et al. monitored alertness and wakefulness in their participants, although the former study notes that participants did not report drowsiness. It is a limitation of our study as well that drowsiness and alertness were not quantitatively monitored throughout, either using EEG-based metrics or rating scales. However, they were qualitatively monitored throughout the study, as in similar studies[18].

Our findings are in line with studies employing similar methods[18] and coil shapes and sizes[17] and suggest that there is no effect of brief visual deprivation on motor cortex excitability in awake individuals. As such, closing of eyes for brief periods may improve relaxation or comfort of study participants without compromising comparability to measures recorded with eyes open, provided alertness is maintained.

*Limitations*

Our study recruited 21 healthy volunteers. While our sample size is similar to previous studies, for some conditions RMT data was missing for 3 and SICI was missing for 8 participants. Smaller sample size may have affected the sensitivity of our measures. While most participants in our study were young (18-27 years old), one outlier was a 50-year-old participant. Excluding this participant from our analyses did not alter results. While no significant age difference was found between sexes, effects of sex were not considered in our models. As the aim of this study was to investigate the lurking variables of eye state and coil size across the population, sex-balanced recruitment was undertaken to ensure equal representation of males and females. Larger sample sizes would be warranted to simultaneously interrogate the effect of sex on all factors of interest examined in this study.

Finally, drowsiness and alertness were not quantitatively monitored throughout, either using EEG-based metrics or rating scales. However, they were qualitatively monitored throughout the study.

**Conclusion**

We found that RMT significantly differs when using a small 30BFT coil but not when using standard 50BFT or 70BF coils. Coil size had no effect on SICI. Likewise, there were no significant effects of eye state on RMT or SICI, including when controlling for coil size. Our findings have implications for future TMS-based studies, and for reliability assurance in large multicenter studies. We suggest that the depth-focality trade-off and participant comfort levels with respect to coil size should be considered. As 50BF and 70BF coils appear to produce similar single and paired pulse results, their variable use across research centers may not impact reliability. While participants are typically instructed to keep their eyes open during TMS studies, our findings suggest that brief periods of eye closure in the absence of sleep do not affect cortical excitability.

**Financial support:** This research did not receive any specific grant from funding agencies in the public, commercial, or not-for-profit sectors

**Authorship statement:** Roisin McMackin and Narin Suleyman conceptualized the study. Roisin McMackin and Gabriel Palma designed the study methodology, generated codes and figures, Meher Sabharwal, Roisin McMackin and Gabriel Palma performed data analysis. Narin Suleyman and Meher Sabharwal undertook data collection and generated the original manuscript draft. Roisin McMackin supervised the project. All authors reviewed and edited the manuscript and approved of the final submission.

**Conflict of interest statement:** Dr McMackin declares consultancy fees from The Science Behind Ltd. and honoraria from Brainbox Ltd.